\documentclass[twocolumn,twoside,11pt]{elsart}
\usepackage{graphicx}
\usepackage{epsfig}
\usepackage{amssymb}
\def\aA{$\alpha$-nucleus\ }
\def\AA{nucleus-nucleus\ }

\def\oo{$^{16}$O+$^{16}$O\ }
\def\bm#1{\mbox{\boldmath $#1$}}

\def\pA{proton-nucleus\ }
\def\phe6{$^6$He+$p$\ }
\def\he6pn{$p(^6$He,$^6$Li$^*)n$\ }

\def\pphe6{$p(^6$He,$^6$He)$p$\ }
\def\p2he6{$p(^6$He,$^6$He$^*)p$'\ }
\def\nli6{$^6$Li$^*+n$\ }
\def\Ca48pn{$^{48}$Ca$(p,n)^{48}$Sc\ }

\begin{document}
\begin{frontmatter}
\title{Probing the Nuclear Equation of State in the quasi-elastic
 nucleus-nucleus scattering}
\author[INST]{Dao T. Khoa},
\author[HMI]{W. von Oertzen},
\author[HMI]{H.G. Bohlen}
and \author[INST]{H.S. Than}
\address[INST]{Institute for Nuclear Science and
 Technique, P.O. Box 5T-160, Nghia Do, Hanoi, Vietnam.}
\address[HMI]{Hahn-Meitner-Institut Berlin GmbH, Glienicker Str. 100,
D-14109 Berlin, Germany.}
\begin{abstract}\rm
Large-angle elastic scattering of $\alpha$-particle and strongly-bound light
nuclei at a few tens MeV/nucleon has shown the pattern of \emph{rainbow
scattering}. This interesting process was shown to involve a significant overlap
of the two colliding nuclei, with the total nuclear density well above the
saturation density of normal nuclear matter (NM). For a microscopic calculation
of the nucleus-nucleus potential within the folding model, we have developed a
\emph{density dependent} nucleon-nucleon (NN) interaction based on the G-matrix
interaction M3Y. Our folding analysis of the refractive $^4$He, $^{12}$C, and
$^{16}$O elastic scattering shows consistently that the NM incompressibility $K$
should be around 250 MeV which implies a rather \emph{soft} nuclear Equation of
State (EOS). To probe the symmetry part of the nuclear EOS, we have used the
isovector coupling to link the isospin dependence of the proton optical
potential to the cross section of $(p,n)$ charge-exchange reactions exciting the
isobaric analog states in nuclei of different mass regions. With the isospin
dependence of the NN interaction fine tuned to reproduce the charge exchange
data, a realistic estimate of the NM symmetry energy has been made.
\end{abstract}
\end{frontmatter}

\section{What is the nuclear rainbow?}
The atmospheric rainbow is observed in nature whenever there are water droplets
illuminated by the sun light. It can be seen during the rain with the sunshine
not completely covered by the clouds or from a fountain when the sunlight enters
from behind the point of observation. Beside the fascinating effect of color
splitting in the rainbow caused by the dependence of the refraction index on the
light wavelength, a physically more interesting effect is the \emph{increased}
light intensity around the rainbow angle $\Theta_R$ and the \emph{shadow} region
lying beyond $\Theta_R$ which are results of a particular refraction -
reflection sequence.

Although Descartes has successfully explained the origin of the atmospheric
rainbow based on simple geometrical ray optics in 1637, neither he nor Newton
and Young (several decades later) could explain the fine structure of the
rainbow seen as the \emph{supernumeraries} (the faint bows located just below
the primary bow). The first complete mathematical description of the atmospheric
rainbow was given in 1838 by Airy, and the oscillation of the light intensity
near the rainbow angle is now known as Airy oscillation which gives rise to the
supernumeraries. As nuclei are approximately spherical objects (like water
drops) having wave properties, they can be refracted or undergo interference in
the \AA scattering, like the refraction of sunlight by the water drops. As a
result, one may observe phenomena like rainbow in the nuclear scattering if the
conditions are right. Indeed, the rainbow pattern was clearly observed in
large-angle elastic scattering of $\alpha$-particles and some strongly-bound
light nuclei at few tens MeV/nucleon, where signatures of the Airy oscillation
pattern has been identified. An important feature of the nuclear rainbow is that
it allows us to probe the \AA interaction at small distances \cite{vOe00} thanks
to a \emph{weak} absorption in the \AA system.

\section{From nuclear rainbow to the equation of state for cold nuclear matter}
\begin{figure}[htb]
 \vspace*{-1.3cm}\hspace*{-0.5cm}
 \mbox{\epsfig{file=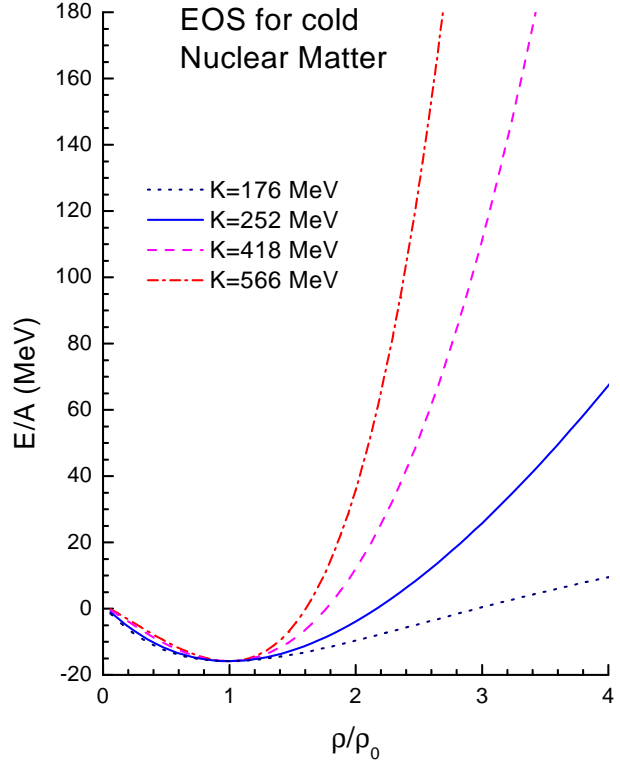,height=12.6cm}}\vspace*{-1cm}
\caption{EOS for cold NM given by the HF calculation using different density
dependent NN interactions (which give different values of the NM
incompressibility $K$).}\label{f01}
\end{figure}

For a microscopic description of the elastic \AA scattering, the folding model
analysis is usually performed, where the (real) \AA optical potential is
calculated as a Hartree-Fock (HF) potential of the dinuclear system
\cite{Kho94,Bra97} using an effective nucleon-nucleon (NN) interaction
\cite{Kho93}
\begin{equation}
 V=\sum_{ij}[<ij|v_D|ij>+<ij|v_{EX}|ji>],
\label{ef1}
\end{equation}
where $|i>$ and $|j>$ are the single-particle wave functions of nucleons in the
two colliding nuclei $A_1$ and $A_2$, respectively; $v_{D}$ and $v_{EX}$ are the
direct and exchange parts of the effective NN interaction. It turns out that the
density dependence of the effective NN interaction can be accurately tested in
the folding analysis of the refractive elastic \aA or \AA scattering data
\cite{Kho94,Kho95}. For this purpose, a phenomenological density dependence was
first introduced to the M3Y interaction based on the G-matrix elements of the
Reid and Paris NN potentials \cite{Be77}, to reproduce to the saturation
properties of cold nuclear matter (NM) in the HF scheme \cite{Kho93}. From the
HF results for the NM energy (per nucleon) plotted in Fig.~\ref{f01}, one can
see that different sets of the density dependence give values of the NM {\em
incompressibility} $K$ ranging from 170 to above 500 MeV. Since the $K$ value is
a key input in the NM equation of state (EOS), a test of the density dependence
of the NN interaction is also an indirect test of the nuclear EOS.
In the elastic channel, the two colliding nuclei remain in the ground state
(g.s.) even when they overlap strongly at small impact parameters, since any
density deformation or rearrangement directly projects the system out of the
elastic channel. Therefore, the total density $\rho$ of the two overlapping
nuclei which enters Eq.~(\ref{ef1}) must be taken as the sum of the two g.s.
densities and the total density for a projectile overlapping a target nucleus
may reach as much as twice the normal NM density $\rho_0$ \cite{vOe00,Kho94}.
\begin{figure}[htb]
 \vspace*{-2cm}\hspace*{-1cm}
 \mbox{\epsfig{file=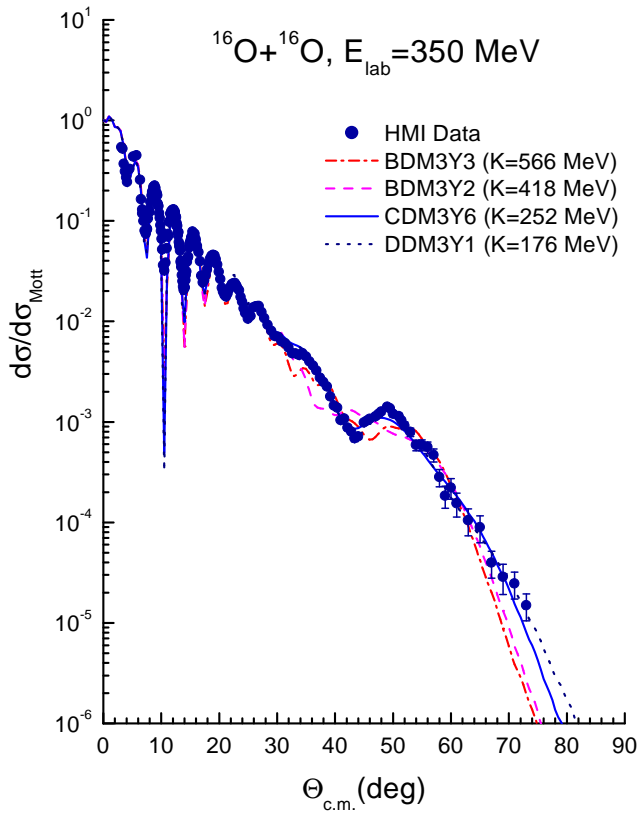,height=13cm}}\vspace*{-1cm}
\caption{Folding model description of the elastic \oo scattering data at
$E_{\rm lab}=350$ MeV \cite{Sti89} using the same density dependent NN
interactions as those in Fig.~\ref{f01}. The best-fit interaction is CDM3Y6
which gives $K\approx 252$ MeV.}\label{f02}
\end{figure}
Note that the shorter the impact parameter the higher the overlap density and
the more the \AA potential becomes sensitive to the density dependence of the NN
interaction. At small impact parameters (or low partial waves) the survival
probability of the elastic wave is usually less than 1\% and requires,
therefore, a very precise measurement of elastic events at large scattering
angles. These large-angle data points give us the most vital information about
the reliability of the effective NN interaction used in the folding calculation
\cite{Kho97}. From results of the folding analysis of the elastic \oo scattering
data at 350 MeV \cite{Sti89} plotted in Fig.~\ref{f02} one can clearly see which
is the most appropriate density dependence of the NN interaction (in the
corresponding $K$ value). Together with the results obtained earlier for elastic
scattering of $\alpha$ particles and other light projectiles \cite{Kho97}, we
conclude that the most realistic value for the NM incompressibility $K$ is
around 230 - 270 MeV which corresponds to a rather \emph{soft} EOS.
\begin{figure}[htb]
 \vspace*{-2cm}\hspace*{-1cm}
 \mbox{\epsfig{file=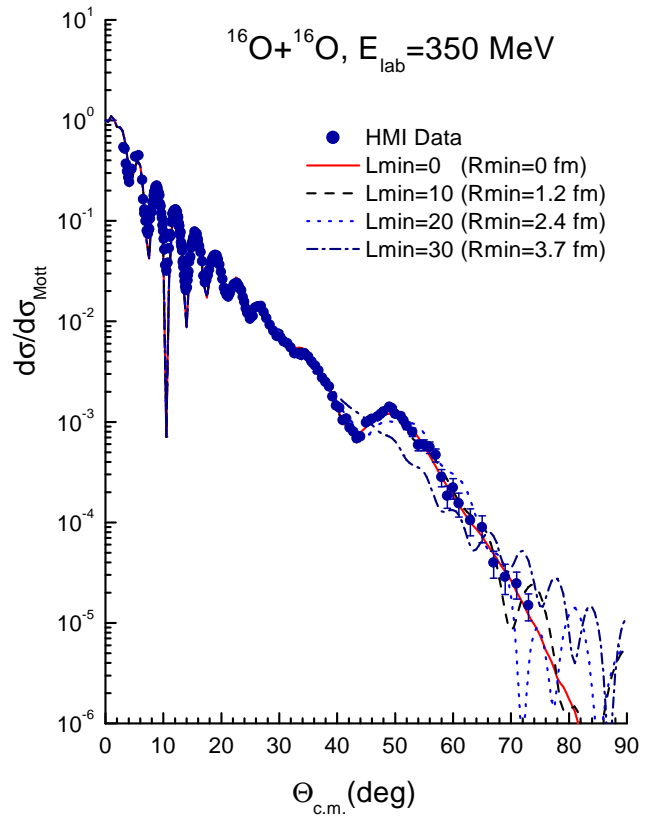,height=13cm}}\vspace*{-1cm}
\caption{The same description as in Fig.~\ref{f02} given by the CDM3Y6
interaction, but with different cutoff values of the lowest partial wave
 L$_{\rm min}$ (and the corresponding impact parameters
 R$_{\rm min}$).}\label{f03}
\end{figure}
We stress that the broad bump seen in Fig.~\ref{f02} at $\Theta_{\rm
c.m.}\approx 50^\circ$ has been specified as the first Airy maximum
\cite{Kho00}, and the observed rainbow pattern is quite sensitive to the \oo
optical potential at small distances. It can be shown for the considered \oo
system (see Fig.~\ref{f03}) that the large-angle data are sensitive to the
impact parameters as small as 2 fm. At such small internuclear distances, the
total overlap density of the system reaches up to 2$\rho_0$ \cite{Kho97}.
Therefore, the elastic refractive \AA scattering data like those measured for
the \oo system at 350 MeV provide a very good data base for probing the high
overlap density in the elastic channel.

\section{Probing EOS of the asymmetric NM via charge-exchange reaction}
The nuclear EOS presented in Fig.~\ref{f01} was obtained in a HF calculation for
symmetric NM, i.e., with equal proton and neutron densities. In reality, the NM
that exists in the neutron stars is highly asymmetric, with the neutrons
outnumbering protons by factor of about 2 in the crust of a neutron star.
Therefore, the knowledge about the symmetry part of the EOS is vital for the
understanding of the dynamics of supernovae explosion and the formation of
neutron stars \cite{Bet90,Swe94}. The symmetry part of the nuclear EOS is
determined essentially by the NM symmetry energy $S(\rho)$ (defined in terms of
a Taylor series expansion of the NM binding energy $B\equiv E/A)$ as
\begin{equation}
 B(\rho,\delta)=B(\rho,0)+S(\rho)\delta^2+O(\delta^4)+...
 \label{e1}
\end{equation}
where $\delta=(\rho_n-\rho_p)/\rho$ is the neutron-proton asymmetry parameter.
The contribution of $O(\delta^4)$ and higher-order terms in Eq.~(\ref{e1}),
i.e., the deviation from the parabolic law was proven to be negligible
\cite{Kho96}. The NM symmetry energy determined at the NM saturation density,
$E_{\rm sym}=S(\rho_0)$ with $\rho_0\approx 0.17$ fm$^{-3}$, is widely known in
the literature as the \emph{symmetry energy} or symmetry coefficient. Although
numerous nuclear many-body calculations have predicted $E_{\rm sym}$ to be
around 30 MeV, a direct experimental determination of $E_{\rm sym}$ still
remains a challenging task. One needs, therefore, to relate $E_{\rm sym}$ to
some experimentally inferrable quantity like the neutron skin in neutron-rich
nuclei \cite{Fur02} or the fragmentation data of heavy-ion (HI) collisions
involving $N\neq Z$ nuclei \cite{Ono03}. Within the frame of any microscopic
model for asymmetric NM, the symmetry energy depends strongly on the isospin
dependence of the NN interaction used therein \cite{Kho96}. Therefore, the
$E_{\rm sym}$ value can be indirectly tested in a charge exchange (isospin-flip)
reaction which has been known for decades as a good probe of the isospin
dependence of the effective NN interaction \cite{Doe75}. Although the isospin
dependence of the nuclear optical potential (OP), known by now as Lane potential
\cite{La62}, has been studied since a long time, there has been a considerable
interest recently in studying the isospin dependence of the OP in the
quasi-elastic scattering reactions measured with unstable neutron-rich beams.
Based on the isospin symmetry, the \AA OP can be written in terms of an
isovector coupling \cite{La62} as
\begin{equation}
 U(R)=U_0(R)+4U_1(R)\frac{{\bm t}.{\bm T}}{aA}, \label{e2}
\end{equation}
where ${\bm t}$ is the isospin of the projectile $a$ and ${\bm T}$ is that of
the target $A$. While the relative contribution by the Lane potential $U_1$ to
the elastic ($p,p$) cross section is small and amounts only to a few percent for
a neutron-rich target, it determines entirely the (Fermi-type) $\Delta
J^\pi=0^+$ transition strength of the ($p,n$) reaction leading to an isobaric
analog state (IAS). Therefore, the ($p,n$) reaction has been so far the main
tool in studying the isospin dependence of the \pA OP. Since this isospin
dependence should be better tested in the charge exchange reactions induced by
the neutron-rich beams, we consider in the present work the \he6pn reaction
measured by Cortina-Gil {\sl et al.} \cite{Gil98} with the secondary $^6$He beam
at $E_{\rm lab}=41.6A$ MeV. Given a large neutron-proton asymmetry
($\delta=1/3$) of the unstable $^6$He nucleus, the measured \he6pn cross section
for the transition connecting the ground state of $^6$He ($T=T_z=1$) and its
isobaric analog partner ($T=1, T_z=0, J^\pi=0^+$ excited state of $^6$Li at
3.563 MeV) is indeed a good probe of the isovector coupling in the \phe6 system.
To link the Lane potential $U_1$ to the isospin dependence of the $NN$
interaction, we have used the folding model \cite{Kho02} to calculate $U_0$ and
$U_1$ using the explicit proton and neutron g.s. densities of $^6$He and the
CDM3Y6 density- and isospin dependent NN interaction \cite{Kho97}. The only
nuclear structure input is the $^6$He$_{\rm g.s.}$ density and we have used the
microscopic density given by the cluster-orbital shell model approximation
(COSMA) \cite{Kor97}.
\begin{figure}[htb]
 \mbox{\epsfig{file=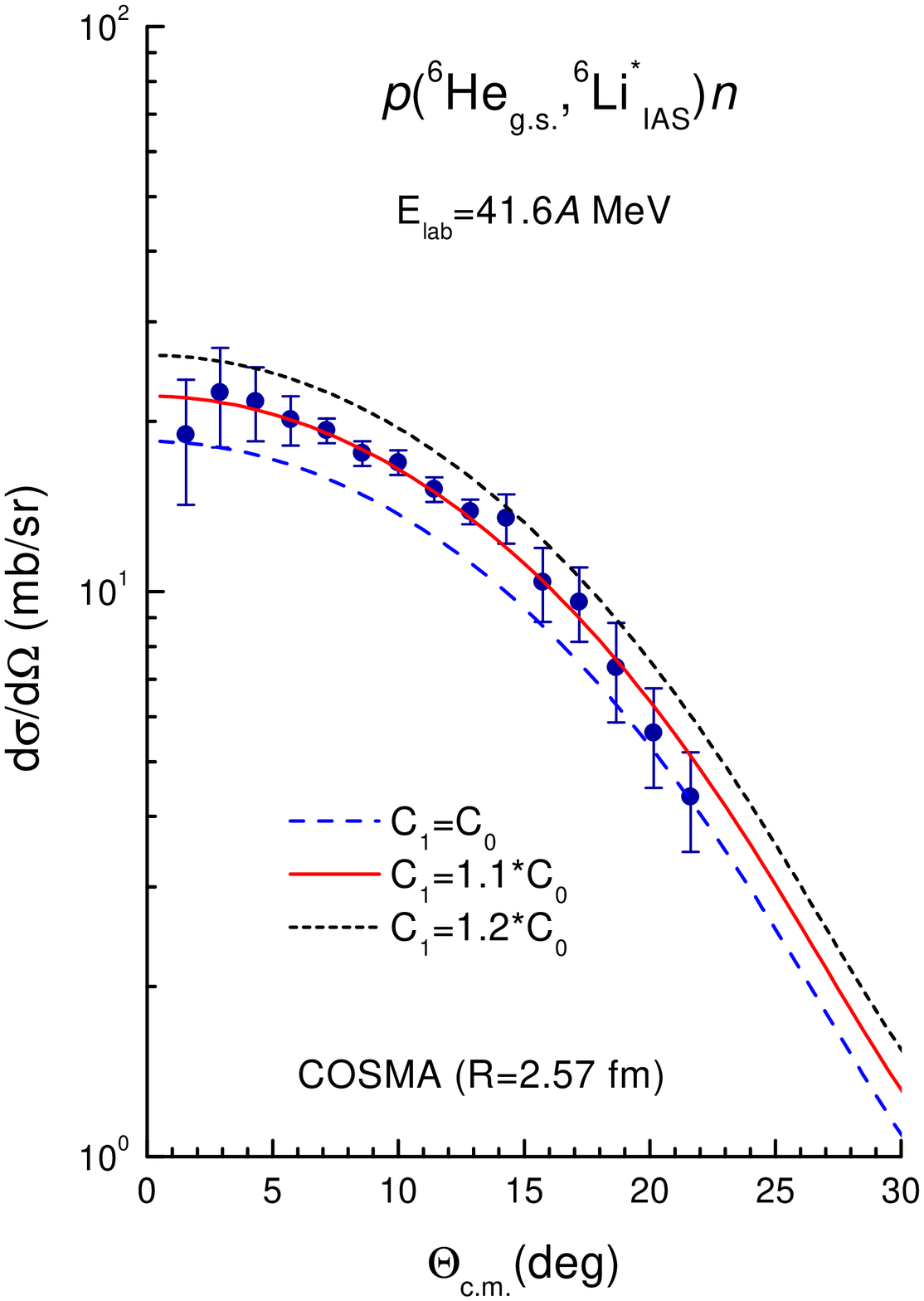,height=12cm}}\vspace*{-1cm}
\caption{CC results for the charge exchange \he6pn cross section at $E_{\rm
lab}=41.6A$ MeV in comparison with the data measured by Cortina-Gil {\sl et
al.} \cite{Gil98}.} \label{f1}
\end{figure}
The \he6pn cross sections given by the coupled-channel (CC) calculation using a
charge-exchange form factor based on the Lane potential $U_1$ were found to have
a shape very close to that of the measured angular distribution (see
Fig.~\ref{f1}). Since the complex strength of the form factor was fixed by the
folding model analysis of the elastic \phe6 scattering \cite{Kho05}, the CC
description of the \he6pn data could be improved only by fine tuning the
strength $C_1$ of the isovector part of the density dependence of the CDM3Y6
interaction \cite{Kho96}. One can see that the best fit is achieved when $C_1$
is about 10\% stronger than the isoscalar strength $C_0$.
\begin{figure}[htb]
 \vspace*{-1.5cm}\hspace*{-0.8cm}
 \mbox{\epsfig{file=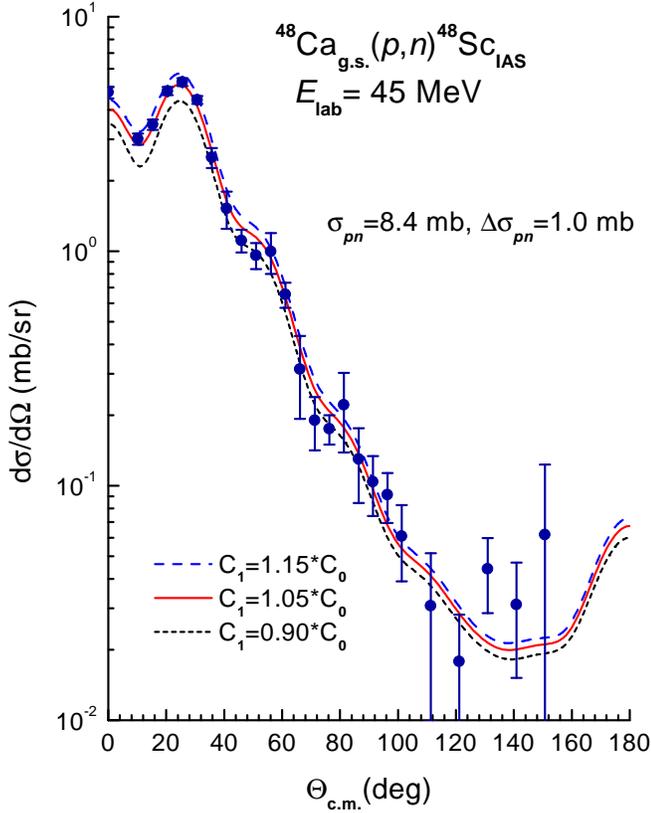,height=13cm}}\vspace*{-1cm}
\caption{CC results for the charge exchange \Ca48pn cross section at the
incident proton energy of 45 MeV in comparison with the data measured by Doering
{\sl et al.} \cite{Doe75}.} \label{f2}
\end{figure}
To make a more definitive conclusion on the EOS of asymmetric NM, we found it
necessary to make a systematic folding analysis of the charge exchange $(p,n)$
reactions measured with targets in different mass regions. Therefore, the same
density- and isospin dependent NN interaction has been used to construct the
charge exchange form factors for the $(p,n)$ reactions measured at the incident
proton energies of 35 and 45 MeV with the targets $^{48}$Ca, $^{90}$Zr,
$^{120}$Sn and $^{208}$Pb \cite{Doe75}. Although the neutron-proton asymmetry
$\delta$ of these nuclei is smaller than that of unstable $^6$He nucleus, the
complex proton OP for these nuclei has been studied over the years \cite{CH89}
and, hence, it allows us to unambiguously probe the isovector part of the OP via
the $(p,n)$ reaction. By using a most appropriate complex proton OP which
produces not only the elastic scattering data but also the polarization data and
the experimental total reaction cross section for each of the considered
targets, we have come up with about the same accurate description of the $(p,n)$
reactions leading to the excitation of IAS in $^{48}$Ca, $^{90}$Zr, $^{120}$Sn
and $^{208}$Pb as that presented above for the unstable $^{6}$He. Our CC results
for the charge exchange \Ca48pn cross section at 45 MeV and the data measured by
Doering {\sl et al.} \cite{Doe75} are shown in Fig.~\ref{f2} as an illustration
of the success of our approach. The best-fit isovector strength $C_1$ of the
CDM3Y6 interaction was adjusted in each case not only to reproduce the measured
angular distribution for the considered $(p,n)$ reaction but also to obtain the
measured \emph{total} $(p,n)$ cross section in our CC calculation. For example,
the $C_1$ values used to calculate the \Ca48pn cross section shown in
Fig.~\ref{f2} was constrained by a CC calculation giving the experimental total
\Ca48pn cross section of $8.4\pm 1.0$ mb \cite{Doe75}.
\begin{figure}[htb]
 \hspace*{-0.7cm}
 \mbox{\epsfig{file=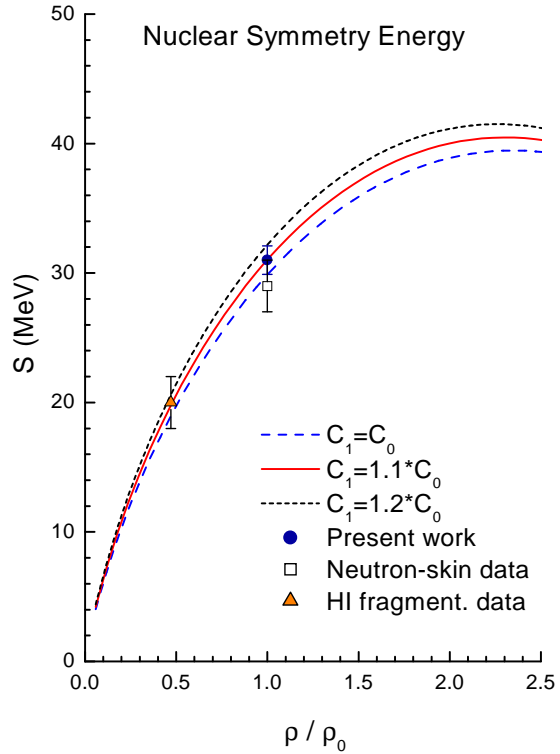,height=12cm}}\vspace*{-1cm}
\caption{Density dependence of the NM symmetry energy $S(\rho)$ predicted by the
HF formalism \cite{Kho96} using the same isovector strengths $C_1$ as those used
in Fig.~\ref{f1} and the (empirical) neutron-skin \cite{Fur02} and HI
fragmentation \cite{Ono03} data.} \label{f3}
\end{figure}
If one takes into account only the uncertainty of the measured $(p,n)$ angular
distribution, the range of acceptance for $C_1$ values becomes significantly
larger. Given the results of our CC calculation of the $(p,n)$ reactions for
$^{48}$Ca, $^{90}$Zr, $^{120}$Sn and $^{208}$Pb \cite{Kho05a}, we conclude that
the best-fit isovector strength $C_1$ of the CDM3Y6 interaction is slightly
larger than the isoscalar strength $C_0$, with about the same uncertainty as
that found in the analysis of \he6pn reaction.

With the isovector strength of the CDM3Y6 interaction now well tested, we have
further performed the HF calculation \cite{Kho96} of asymmetric NM using this
same isospin- and density dependent interaction. The density dependence of the
NM symmetry energy $S(\rho)$ obtained with the same isovector strengths $C_1$ as
those used in Fig.~\ref{f1} is shown in Fig.~\ref{f3}, and one can deduce easily
$E_{\rm sym}\approx 31\pm 1$ MeV from our HF results. This result is quite
complementary to the structure studies which relate the $E_{\rm sym}$ value to
the neutron skin, a method first suggested by Brown \cite{Bro00}. If one adopts
a neutron-skin $\Delta R\approx 0.1-0.2$ fm for $^{208}$Pb then a systematics
based on the mean-field calculations \cite{Fur02} gives $E_{\rm sym}\approx
27-31$ MeV (which is plotted in Fig.~\ref{f3}). Our result is also complementary
to the recent studies of HI fragmentation based on the antisymmetrized molecular
dynamics \cite{Ono03} which obtained $S(\rho\approx 0.08$ fm$^{-3})\approx
18-22$ MeV at a \emph{finite} temperature around 3 MeV. If we neglect the
temperature dependence of $S(\rho)$ at low NM densities, this value turns out to
agree well with our HF result for the low-density part of $S(\rho)$ as shown in
Fig.~\ref{f3}.

In conclusion, the charge exchange reaction like $(p,n)$ or $(^3$He,$t)$ should
be a very powerful tool in studying not only the nuclear structure but also the
isospin aspects of the EOS for asymmetric NM.

\section{Acknowledgement}
This work has been supported, in part, by the Natural Science Council of
Vietnam, A.v. Humboldt Stiftung of Germany, EU Asia-Link Program
CN/Asia-Link/008 (94791) and Vietnam Atomic Energy Commission.

\end{document}